\documentclass [12pt]{article}
\usepackage{graphicx,amssymb,amsmath,bm}
\usepackage[cp1251]{inputenc}
\usepackage[english]{babel}
\usepackage[symbol*]{footmisc}

\makeatletter
\def\@biblabel#1{#1.\hskip-0.3em}
\makeatother

\begin{document}
\def\refname{\normalsize \centering References}
\def\abstractname{}

\begin{center}
{\large \bf Microscopic Description of Diffractive Deuteron Breakup\\ by $^{3}\text{He}$ Nuclei}
\end{center}

\begin{center}
\bf\text{V.~I.~Kovalchuk}\footnote{E-mail: sabkiev@gmail.com}
\end{center}

\begin{center}
\small
\textit{Department of Physics, Taras Shevchenko National University, Kiev 01033, Ukraine}
\end{center}

\begin{abstract}
A microscopic formalism for describing observed cross sections for deuteron breakup by
three-nucleon nuclei was developed on the basis of the diffraction nuclear model.
A general formula that describes the amplitude for the reaction
$^{2}\text{H}(^{3}\text{He},^{3}\text{He}\,p)n$ and which involves only one adjustable
parameter was obtained by using expansions of the integrands involved in terms of a
Gaussian basis. This formula was used to analyze experimental data on the exclusive cross
sections for deuteron breakup by $^{3}\text{He}$ nuclei at the projectile energy
of 89.4~MeV. The importance of employing, in calculations, a deuteron wave function
that has a correct asymptotic behavior at large nucleon-nucleon distances was demonstrated.

\end{abstract}

\bigskip
\begin{center}
\bf 1. Introduction
\end{center}
\smallskip

Investigations between collisions of light nuclei treated as composite particles are an important source
of information about the microscopic structure of these nuclei and about mechanisms of nuclear reactions
proceeding under specific kinematical conditions. While one can still describe reactions involving three
or four particles on the basis of a rigorous theory (for example, by invoking, respectively, the formalism
of Faddeev equations or the formalism of Faddeev-Yakubovsky equations), a description of the interaction
between nuclear systems featuring a greater number of particles already requires employing approximate
methods.

The diffraction model of multiparticle collisions~\cite{1}, which admits the inclusion of a microscopic
description of nucleon-density distributions, nucleon-nucleon phase shifts, and the corresponding
profile functions, can be considered as one such approach. This permits minimizing the number of adjustable
parameters, whereby one can obtain a quantitative description of relevant experimental data and verify
the model itself and its applicability boundaries for the kinematics of the reaction being considered.
By and large, the application of this approach complicates somewhat the formalism used: for example,
the integrands in the scattering amplitude develop a dependence on multiple integrals. As will be shown below,
however, the ultimate expression for the reaction amplitude can be reduced to an algebraic expression -- multiple
sums involving elementary functions -- if Gaussian functions are used as integrands. Here, Gaussian functions
can be used as basis functions in expansions of both wave functions for colliding nuclei (variational problem)
and arbitrary profile functions. It is noteworthy that a similar procedure was extensively employed in the
variational approach in describing bound states of light nuclei~\cite{2,3,4}, in parametrizations of ground-state
charge densities of nuclei~\cite{5,6}, and in the problems of scattering~\cite{7} and deuteron stripping~\cite{8}.
This makes it possible to calculate analytically respective phase shifts and form factors.

We note that, in~\cite{9}, the diffraction model of multiple scattering was already used to describe
the breakup reaction $^{3}\text{H}(d,p)n{^{3}}\text{H}$. The approach proposed in [9] is formally microscopic
because it is based on nucleon-nucleon interactions, but the authors of that study used profile functions
including free parameters and replaced the deuteron wave function by a simple Gaussian function,
which do not reproduce the asymptotic behavior of the nuclear density at large nucleon-nucleon distances.
Since the cross sections for deuteron-breakup reactions (in just the same way as those for deuteron-stripping
reactions) are sensitive to the behavior of the deuteron wave function in the region of its tail~\cite{10},
one can consider the results obtained in~\cite{9} as quite approximate ones.

In the present study, exclusive spectra for deuteron breakup by $^{3}\text{He}$ nuclei at the energy
of 89.4 MeV~\cite{11} were chosen as the object of our analysis based on the microscopic diffraction
nuclear model. Here, we do not take into account Coulomb interaction because, in our case, the barrier
height is approximately equal to 1.5~MeV, so that the emission angles of particles exceed $20^{\circ}$.
We do not take into account particle spins either.

\bigskip
\begin{center}
{\bf 2. Formalism}
\end{center}
\smallskip

Let us introduce the coordinate frame in which the coordinate origin coincides with the center of mass
of the target nucleus (deuteron), while the positive direction of the $z$-axis coincides in direction with
the projectile momentum ${\bf k}$ (see Fig.~1).
\vspace{3mm}
\begin{figure}[!h]
\center
\includegraphics [scale=0.52] {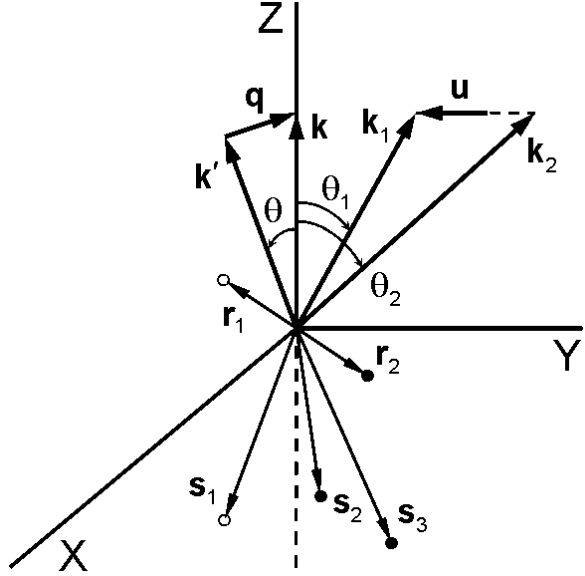}
\caption{Coordinate frame used in calculating the amplitude for deuteron breakup by ${^3}\text{He}$
nuclei. The open and closed circles represent neutrons and protons, respectively.}
\label{Fig1}
\end{figure}
\vspace{3mm}

In this coordinate frame, the amplitude for diffractive deuteron breakup by ${^3}\text{He}$ nuclei
can be represented, according to~\cite{1}, in the form
\begin{equation}
F({\bf q, u})=\frac{ik}{2\pi}\int d{\bf b}\exp(i{\bf qb})\int d{\bm\tau}
\Phi_{\bf u}^{*}(\bm\tau)\sum_{\ell=1}^{3}\Bigl(1-\prod_{j=1}^{2}
\bigl[1-\omega_{j\ell}(b_{j\ell})\bigr]\Bigr)\Phi_{0}(\bm\tau),
\label{eq1}
\end{equation}
where ${{\bf q}={\bf k}-{\bf k}'}$ is the momentum transfer; $k(k')$ is the incident-nucleus (scattered nucleus)
momentum; ${{\bf u}=({\bf k}_1-{\bf k}_2)/2}$ is the relative momentum deuteron-breakup products;
${\bf b}=({\bf s}_1+{\bf s}_2+{\bf s}_3)_{\perp}/3$ is the projection of the vector of the ${^3}\text{He}$-nucleus
center of mass onto the $xy$-plane; $b_{j\ell}=|{\bf b}_{j\ell}|$
(where ${\bf b}_{j\ell}=({\bf r}_j-{\bf s}_{\ell})_{\perp}$ is the impact-parameter vector lying in the
$xy$-plane and corresponding to a collision between the $j{\text{th}}$ deuteron nucleon and the $\ell{\text{th}}$
nucleon of the ${^3}\text{He}$ nucleus); and $\bm\tau$ is the set of coordinates that describe the intrinsic
degrees of freedom of nucleons belonging to the deuteron and ${^3}\text{He}$ nucleus involved.
Further, the quantities $\Phi_0$ and $\Phi_{\bf u}$ in expression (\ref{eq1}) are the wave functions that
describe the system under study, respectively, before and after the collision event; that is,
\begin{equation}
\Phi_{0,{\bf u}}(\bm\tau)=\psi_0({\bf s})\varphi_{0,{\bf u}}(\bf r),
\label{eq2}
\end{equation}
where $\psi_0({\bf s})$ is the ground-state wave function for the ${^3}\text{He}$ nucleus, $\varphi_{0}(\bf r)$
is the ground-state deuteron wave function, and the function $\varphi_{{\bf u}}(\bf r)$ describes the
relative motion of the neutron and proton that originated from deuteron breakup.

Since $d{\bf b}=d^{(2)}{\bf b}$, $d{\bm\tau}=d^{(3)}{\bf s}\,d^{(3)}{\bf r}$, expression (\ref{eq1}) is formally
an eight-dimensi-onal integral. A direct calculation of such integrals presents a difficult challenge from the
point of view of convergence of quadrature sums and computational errors, but, upon expanding the integrands
in terms of a Gaussian basis, the amplitude in (\ref{eq1}) admits an analytic evaluation.

We now recast expression (\ref{eq1}) into the form
\begin{equation}
F({\bf q, u})=\frac{ik}{2\pi}\int d{\bf b}\exp(i{\bf qb})\int d{\bf s}\,d{\bf r}\,
\rho(s)\varphi_{0}({\bf r})\varphi_{{\bf u}}^{*}({\bf r})\sum_{\ell=1}^{3}\Bigl(1-\prod_{j=1}^{2}
\bigl[1-\omega_{j\ell}(b_{j\ell})\bigr]\Bigr),
\label{eq3}
\end{equation}
where $\rho(s)=\psi_0({\bf s})\psi_0^{*}({\bf s})$ is the normalized (to unity) nucleon-density distribution in
the ${^3}\text{He}$ nucleus and $s^2=({\bf s}_1-{\bf s}_2)^2+({\bf s}_2-{\bf s}_3)^2+({\bf s}_1-{\bf s}_3)^2$.
In order to evaluate the integrals in expression (\ref{eq3}), we take the integrands $\rho(s)$, $\varphi_{0}(\bf r)$,
and $\varphi_{{\bf u}}({\bf r})$ in the form
\begin{equation}
\rho(s)=\sum_{n=1}^{K}a_{n}\exp(-b_{n}s^2),
\label{eq4}
\end{equation}
\begin{equation}
\varphi_{0}({\bf r})=\sum_{n=1}^{K}c_{n}\exp(-d_{n}r^2),\quad
r^2={\bf r}^2=|{\bf r}_1-{\bf r}_2|^2,
\label{eq5}
\end{equation}
\begin{equation}
\varphi_{{\bf u}}({\bf r})=\exp(i{\bf u}{\bf r})-
\sqrt{8}\exp\Bigl(-\frac{u^2}{4\lambda}-\lambda r^2\Bigr).
\label{eq6}
\end{equation}
The functions in (\ref{eq4}) and (\ref{eq5}) are the solutions that the use of a Gaussian basis
makes it possible to obtain for the variational problems of the bound states of the ${^3}\text{He}$
nucleus and the deuteron for the K2 nucleon-nucleon potential~\cite{3,4}. At $K=10$, the two functions
in question faithfully reproduce the experimental values of the binding energy and root-mean-square
radii of these nuclei and have a correct asymptotic behavior at short and long nucleon–nucleon distances.
In expression (\ref{eq5}), $\lambda$ is the harmonic-oscillator-potential parameter chosen in such a way
as to reproduce the deuteron binding energy in solving the respective Schr{\"o}dinger equation for the
bound state of the neutron and proton. The authors of~\cite{12} indicate that the wave function~(\ref{eq6})
is not a solution of the Schr{\"o}dinger equation for a continuum, nor does it have an asymptotic
behavior at infinity in the form of a converging spherical wave; it does not satisfy the orthonormalization
condition either:
\begin{equation}
\int\varphi_{{\bf u}'}^{*}({\bf r})\varphi_{{\bf u}}({\bf r})d{\bf r}=
(2\pi)^3\delta({\bf u}'-{\bf u}).
\label{eq7}
\end{equation}
However, as was indicated in~\cite{13,14}, the results of calculations for the deuteron-breakup cross sections
exhibit by and large a low sensitivity to the choice of wave functions describing the relative motion of
the neutron and proton in the final S-wave state. In such problems, one usually chooses $\varphi_{{\bf u}}({\bf r})$
in the form (\ref{eq6}) for the sake of convenience, and this sometimes makes it possible to calculate form factors and
reaction amplitudes analytically.

The nucleon–nucleon profile functions $\omega_{j\ell}$ in (\ref{eq3}) were calculated in the high-energy
approximation~\cite{15}; that is,
\begin{equation}
\omega_{j\ell}(b_{j\ell})=1-\exp(-\phi_{j\ell}(b_{j\ell})),
\label{eq8}
\end{equation}
where $\phi_{j\ell}$ is is the phase shift for the scattering of the $j{\text{th}}$ deuteron nucleon by
the $\ell{\text{th}}$ nucleon of the ${^{3}}\text{He}$ nucleus:
\begin{equation}
\phi_{j\ell}(b_{j\ell})=-\frac{1}{v}\int_{-\infty}^{\infty}dz\,
W\Bigl(\sqrt{b_{j\ell}^2+z^2}\Bigr).
\label{eq9}
\end{equation}
Here, $v$ is the speed of the incident nucleon and $W(r)$ is the imaginary part of the
nucleon–nucleon potential.

Within the double-folding model, the eikonal phase shift in (\ref{eq9}) can be calculated analytically~\cite{16}.
The result has the form
\begin{equation}
\phi_{j\ell}(b_{j\ell})=N_W\frac{\pi^2\,\bar\sigma_{j\ell}\,a_j^3\,a_{\ell}^3\,
\rho_j(0)\,\rho_{\ell}(0)}{a_j^2+a_{\ell}^2+r_0^2}
\exp\Bigl[-\frac{b_{j\ell}^2}{a_j^2+a_{\ell}^2+r_0^2}\Bigr],
\label{eq10}
\end{equation}
where $N_W$ is the normalization parameter of the potential $W(r)$ and $\bar\sigma_{j\ell}$ is the isotopically
averaged total cross section for nucleon–nucleon interaction. For various type of colliding nucleons, we have
$\bar\sigma_{j\ell}=\sigma_{nn}=\sigma_{pp}$ or $\bar\sigma_{j\ell}=\sigma_{np}=\sigma_{pn}$.
The cross sections $\sigma_{pp}$ and $\sigma_{pn}$ are described in terms of phenomenological dependences~\cite{16,17}.

Expression (\ref{eq10}) was derived in~\cite{16} under the assumption that the intranuclear-nucleon-density distribution
$\rho_{j,\ell}(r)$ and the nucleon–nucleon interaction amplitude $f_{j\ell}(b)$ are defined in terms of Gaussian
functions as
\begin{equation}
\rho_{j,\ell}(r)=\rho_{j,\ell}(0)\exp(-r^2/a_{j,\ell}^2),\quad
f_{j\ell}(b)=(\pi r_{0}^2)^{-1}\exp(-b^2/r_{0}^2).
\label{eq11}
\end{equation}
Considering that $\rho_{j}(0)\!=\!\rho_{\ell}(0)\!=\!(a\sqrt{\pi})^{-3}$ and
$a_j\!=\!a_{\ell}\!=\!a\!=r_0\!=\!\sqrt{2/3}\,R_{NN}$~\cite{18}, where $R^{2}_{NN}\cong0,65\,\,\text{fm}^{2}$
is the root-mean-square range of nucleon–nucleon interaction, we recast expression (\ref{eq10}) into the form
\begin{equation}
\phi_{j\ell}(b_{j\ell})=N_W\frac{\bar\sigma_{j\ell}}{3\pi a^2}
\exp\Bigl[-\frac{b_{j\ell}^2}{3a^2}\Bigr].
\label{eq12}
\end{equation}
Expression (\ref{eq12}) was directly used to calculate the profile functions (\ref{eq8}), which were thereupon
expanded in terms of a Gaussian basis as
\begin{equation}
\omega_{j\ell}(b_{j\ell})=\sum_{n=1}^{K}\alpha_{n}^{(j\ell)}
\exp(-\beta_{n}^{(j\ell)}b_{j\ell}^2),\quad
j\ell=pp,pn.
\label{eq13}
\end{equation}
Taking into account the nucleon content of colliding nuclei, we spell out the sum in (\ref{eq3}) as
\begin{equation}
\sum_{\ell=1}^{3}\Bigl(1-\prod_{j=1}^{2}\bigl[1-\omega_{j\ell}\bigr]\Bigr)=
-3(\omega_{pp}+\omega_{pn}-\omega_{pp}\,\omega_{pn}).
\label{eq14}
\end{equation}
Evaluating the integral in expression (\ref{eq3}) with the functions (\ref{eq4})–(\ref{eq6}), (\ref{eq13}),
and with allowance for relation (\ref{eq14}), we obtain
\begin{equation}
F({\bf q, u})=-\frac{i\pi^{4}\sqrt{\pi}k}{4\sqrt{3}}
\bigl(\Omega^{(1)}(q)-\Omega^{(2)}(q)\bigr)f({\bf q, u}),
\label{eq15}
\end{equation}
where
\begin{equation}
\Omega^{(1)}(q)=\sum_{i=1}^{K}\left(\frac{\alpha_{i}^{(pp)}}{\beta_{i}^{(pp)}}
\exp\left[-\frac{q^2}{4\beta_{i}^{(pp)}}\right]+
\frac{\alpha_{i}^{(pn)}}{\beta_{i}^{(pn)}}
\exp\left[-\frac{q^2}{4\beta_{i}^{(pn)}}\right]\right),
\label{eq16}
\end{equation}
\begin{equation}
\Omega^{(2)}(q)=\sum_{i=1}^{K}\sum_{j=1}^{K}
\frac{\alpha_{i}^{(pp)}\alpha_{j}^{(pn)}}{\beta_{i}^{(pp)}+\beta_{j}^{(pn)}}
\exp\left[-\frac{q^2}{4\bigl(\beta_{i}^{(pp)}+\beta_{j}^{(pn)}\bigr)}\right],
\label{eq17}
\end{equation}
\begin{equation*}
f({\bf q, u})=\sum_{i=1}^{K}\sum_{j=1}^{K}
\frac{a_{i}}{b_{i}^{3/2}}\frac{c_{j}}{d_{j}^{3/2}}
\exp\left[-\frac{q^2}{18 b_{i}}\right]
\exp\left[-\frac{u_z^2}{4 d_{j}}\right]
\end{equation*}
\begin{equation}
\times\left(\exp\left[-\frac{(u_{\perp}+q/2)^2}{4 d_{j}}\right]+
\exp\left[-\frac{(u_{\perp}-q/2)^2}{4 d_{j}}\right]-
2\exp\left[-\frac{u_{\perp}^2}{4 d_{j}}+
\frac{q^2}{16\bigl(d_{j}+\lambda\bigr)}\right]
\right).
\label{eq18}
\end{equation}
Here, $u^2=u^2_{\perp}+u^2_{z}$, with ${\bf u}_{\perp}$ lying in the impact–parameter $xy$-plane.
Equations (\ref{eq15})–(\ref{eq18}) were obtained for the case where the vectors ${\bf q}$ and ${\bf u}$
are coplanar, which is valid for the geometry of the experiment reported in~\cite{11} and analyzed below.

\bigskip
\begin{center}
{\bf 3. Results of calculations and their comparison with experimental data}
\end{center}
\smallskip

In~\cite{19}, where the problem of deuteron breakup by medium-mass and heavy nuclei was examined, an
expression for the exclusive deuteron-breakup cross section was obtained in the form
\begin{equation}
\sigma(\theta_1,\theta_2,E_2)=\frac{d^{3}\sigma}{d\Omega_1 d\Omega_2 dE_2}=
\frac{m^{5/2}E_1 E_2}{2k^{2}\sqrt{E_1+E_2+\epsilon}}|F({\bf q},{\bf u})|^{2},
\label{eq19}
\end{equation}
where $m$ is the nucleon mass; $E_1$ and $E_2$ are the energies of, respectively, the neutron and the proton
scattered into the respective solid-angle elements $d\Omega_1$ and $d\Omega_2$; and $\epsilon$ is the deuteron
binding energy.

The detecting system used in~\cite{11} recorded a ${^3}\text{He}$ nucleus scattered at an angle $\theta$ and
one deuteron-breakup product (proton), to which the angle $\theta_2$ and the energy $E_2$ correspond.
These three kinematical variables determine the differential cross section $\sigma(\theta,\theta_2,E_2)$
observed in the experiment under discussion. If the neutron (as a deuteron-breakup product) were detected
instead of the ${^3}\text{He}$ scattered nucleus, we would have had the relation
$\sigma(\theta,\theta_2,E_2)=\sigma(\theta_1,\theta_2,E_2)$. Therefore, experimental spectra from~\cite{11}
can be described by expression (\ref{eq19}) under the condition the emission angles $\theta_1$ and $\theta_2$
for the neutron and proton and their respective energies $E_1$ and $E_2$ are known.

The experiment reported in~\cite{11} was performed in coplanar geometry -- that is, the vectors
${\bf k}$, ${\bf k}'$, ${\bf q}$, and ${\bf k}_2$ lay in the the scattering plane (see Fig.~1).
From the momentum-conservation law ${\bf k}={\bf k}'+{\bf k}_1+{\bf k}_2$, it follows that the vector
${\bf k}_1$ also lies in this plane, and so does therefore the vector ${\bf u}=({\bf k}_1-{\bf k}_2)/2$.
The unknown quantities $\theta_1$ and $E_1$ can be found from a set of nonlinear equations that is obtained
by supplementing the momentum-conservation law with the energy-conservation law
\begin{equation}
\left\{ {\begin{array}{*{20}l}
   {q=q_{\perp}=k'\sin{\theta}}=k_1\sin{\theta_1}+k_2\sin{\theta_2},  \\
   {k=k'\cos{\theta}+k_1\cos{\theta_1}+k_2\cos{\theta_2},}  \\
   {E=E'+E_1+E_2+\epsilon.}  \\
\end{array}} \right.
\label{eq20}
\end{equation}
The set of Eqs.~(\ref{eq20}) is solved numerically -- the sought values of $E'$, $E_{1}$, and $\theta_1$
are found for each set of known values of $E$, $\theta$, $E_{2}$, and $\theta_2$.

That detector in~\cite{11} which recorded the ${^3}\text{He}$ scattered nucleus was arranged at a fixed
angle of $\theta=20^\circ$, and emitted protons were detected in the angular range of
$\theta_2=20^\circ\!\div\!60^\circ$. By way of example, Fig.~2 shows the dependences $q(E_2)$ and $u(E_2)$
calculated at an angle of $\theta_2=20^\circ$ (see Fig.~2a) and the values of the minima
$q_{\text{min}}$ and $u_{\text{min}}$ versus the angle $\theta_2$ (Fig.~2b).

\vspace{3mm}
\begin{figure}[!h]
\center
\includegraphics [scale=0.85] {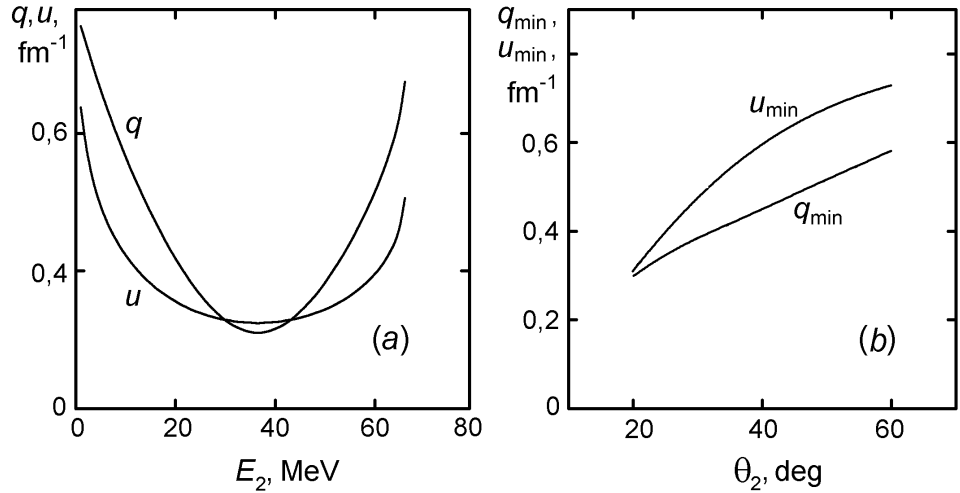}
\caption{Results of solving the set of Eqs.~(\ref{eq20}) at $\theta=20^\circ$:
$(\textit{a})$ $q$ and $u$ versus the proton energy $E_{2}$ at $\theta_2=20^\circ$ and
$(\textit{b})$ values of $q_{\text{min}}$ and $u_{\text{min}}$ versus the angle~$\theta_2$.}
\label{Fig2}
\end{figure}
\vspace{3mm}

This figure shows that the dependences $q(E_2)$ and $u(E_2)$ pass through a minimum at specific values
of the proton energy $E_2$; a similar behavior of the quantities $q$ and $u$ is also characteristic of
the remaining values of the angle $\theta_2$ that were preset in the experiment reported in~\cite{11}.
With increasing $\theta_2$, the proton energy at which $q(E_2)$ and $u(E_2)$ reach a minimum decreases
(not shown in Fig.~2), while the values of $q_{\text{min}}$ and $u_{\text{min}}$ themselves increase.

Figure~3 shows the calculated cross sections~(\ref{eq19}). In this figure, the solid curves were calculated
with the amplitude in~(\ref{eq15}), the normalization parameter of the imaginary part of the high-energy
nucleon–nucleon potential being set to $N_W=0.13$ $(\theta_2=20^{\circ})$; 0.21~$(25^{\circ})$; 0.32~$(30^{\circ})$;
0.53~$(35^{\circ})$; 0.89~$(40^{\circ})$; 0.98~$(45^{\circ})$; 1.0~$(50^{\circ})$; and 1.0~$(55^{\circ})$.

\begin{figure}[!h]
\center
\includegraphics [scale=0.9] {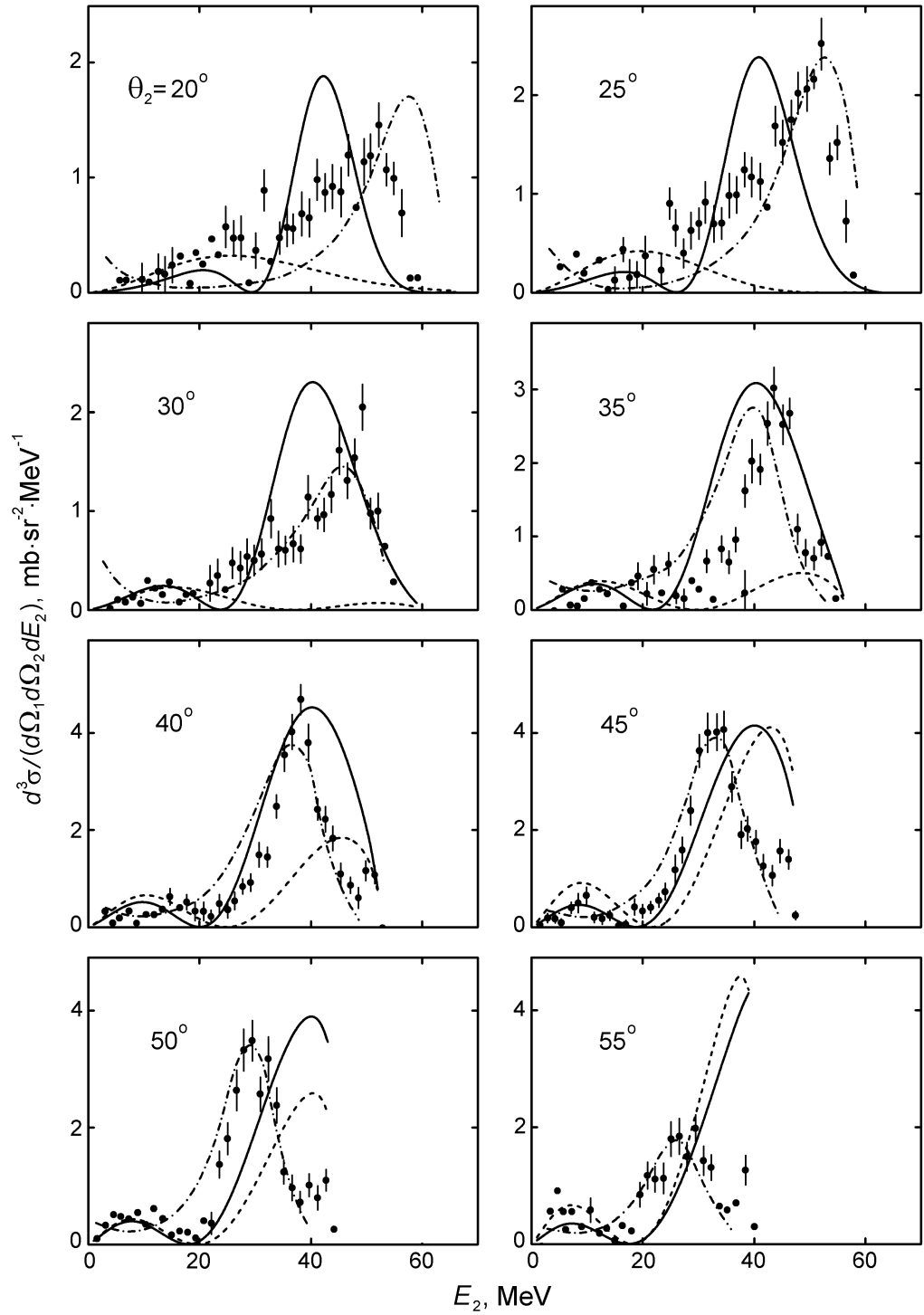}
\caption{Energy distributions of protons from the reaction
\mbox{${^{2}}\text{H}({^{3}}\text{He},{^{3}}\text{He}\,p)n$} induced by ${^3}\text{He}$
projectile nuclei of energy 89.4~MeV. The notation for the curves is explained
in the main body of the text. The displayed experimental data (points) were borrowed
from [11].}
\label{Fig3}
\end{figure}

The dashed curves in Fig.~3 represent the results of the calculations with the simple model deuteron wave
function
\begin{equation}
\varphi_{0}({\bf r})=\left(\frac{2\lambda}{\pi}\right)^{3/4}\exp(-\lambda r^2)
\label{eq21}
\end{equation}
and the nucleon–nucleon profile functions in the form~\cite{9}
\begin{equation}
\omega_{j\ell}(b_{j\ell})=\alpha \exp(-b_{j\ell}^2/R_{NN}),\quad
j\ell=pp,pn,
\label{eq22}
\end{equation}
where $\alpha$ is an adjustable parameter ($\alpha=0.12$ $(\theta_2=20^{\circ})$;
0.18~$(25^{\circ})$; 0.26~$(30^{\circ})$; 0.47~$(35^{\circ})$; 0.73~$(40^{\circ})$; 0.75~$(45^{\circ})$;
0.86~$(50^{\circ})$; 1.0~$(55^{\circ})$), \mbox{$R_{NN}\cong0.65\,\,\text{fm}^{2}$}.
From the behavior of the dashed curves in this figure, it follows that the use of the functions
(\ref{eq21}) and (\ref{eq22}) is a rather rough approximation since this makes it possible to describe
experimental data primarily in the proton-energy range of \mbox{$E_2\le20\div30\,\,\text{MeV}$}.
Such behavior stems first of all from an incorrect asymptotic behavior of the wave function in~(\ref{eq21})
to which the cross section for breakup reactions shows a rather high sensitivity~\cite{10}.

The dash-dotted curves in Fig.~3 were borrowed from~\cite{11}; these cross sections were calculated on the
basis of the four-body scattering model~\cite{20}, whose equations were used to describe deuteron breakup according
to the \mbox{$^3\text{He}+d\to\,p+d+p+n$} scheme with allowance for single collisions exclusively. In~\cite{11},
the absolute cross-section values were also normalized to the respective experimental-peak heights.

It is peculiar to the experimental cross sections from~\cite{11} that they exhibit two peaks -- a small one
in the range of \mbox{$E_2\!=\!10\div20\,\,\text{MeV}$} and a large one in the range of
\mbox{$E_2\!=\!30\div50\,\,\text{MeV}$} (see Fig.~3). From the behavior of the solid curves in Fig.~3,
one can see that the structure of the cross sections calculated on the basis of the formalism outlined in
the present study involves both peaks (in contrast to~\cite{11}, where the first peak was described only
qualitatively, while the second one was described quantitatively in some cases). This feature of the cross
sections in question is determined primarily by the behavior of the parenthetical expression on the
right-hand side of~(\ref{eq18}): it vanishes at \mbox{$E_2\!=\!0$ $(\theta_2\!=\!\theta_1)$}, reaches a maximum
at \mbox{$\theta_1\!=\!0^{\circ}$}, and passes through a minimum at some point $E'_2$; in the vicinity
of this point, we have \mbox{$q(E'_2)\!=\!q_{\text{min}}$} and \mbox{$u(E'_2)\!=\!u_{\text{min}}$}.
As $E_2$ increases further, the parenthetical expression in question begins growing once again, but,
because of the presence of a $q^2$-dependent exponential term in the summand, only up to a specific value
(second maximum).

From a comparison of the solid curves with experimental data from~\cite{11}, it can be seen that the first
maximum is described satisfactorily in the majority of cases (at least, its presence is reproduced).
As for the second maximum, the agreement with experimental data is qualitative here in contrast to the results
from~\cite{11} (dash-dotted curves): with increasing proton emission angle $\theta_2$, the half-width of
the calculated cross section grows, while the position of the peak on the $E_2$ axis remains nearly unchanged.
The reason behind this discrepancy is that, with increasing $\theta_2$, the minimum values $q_{\text{min}}$
and $u_{\text{min}}$ of, respectively, the momentum transfer and the relative momentum become higher (see Fig. 2).
Since the restriction \mbox{$q,u\!<\!<\!1$} is the condition initially imposed for the diffraction approximation
to be applicable to describing breakup reactions, it can be stated that, with increasing $\theta_2$,
this condition is violated.

\bigskip
\begin{center}
{\bf 4. Conclusions}
\end{center}
\smallskip

Within the diffraction nuclear model, a microscopic formalism has been developed for describing
exclusive cross sections for reactions of deuteron breakup by three-nucleon nuclei
(\mbox{$3+2\to3+1+1$} five-body problem). This formalism involves only one adjustable parameter
$N_W$, which is the normalization parameter of the imaginary part of the high-energy nucleon–nucleon
potential.

In deriving the required expression for the reaction amplitude, use was made of expansions of the integrands
involved in terms of a Gaussian basis. This made it possible to evaluate analytically the originally
arising eight-fold integral and to reduce thereby a general expression for the amplitude in question to
triple and quadruple sums.

The resulting formula was used to analyze the experimental exclusive cross section for deuteron
breakup by $^3\text{He}$ nuclei at the projectile energy of 89.4~MeV. The importance of employing,
in calculations, a deuteron wave function that has a correct asymptotic behavior at large nucleon–nucleon
densities was demonstrated in describing experimental data.

The proposed principle for constructing a microscopic formalism for describing the diffractive breakup
reaction can also be employed in other similar few-body problems, such as deuteron breakup
by deuterons, the breakup of tritons and $^3\text{He}$ nuclei by nucleons, and so on.

\end{document}